\documentclass[12pt]{article}

\usepackage{amssymb,fullpage,graphicx}

\newcommand{\ket}[1]{|#1\rangle}
\newcommand{\bra}[1]{\langle#1|}

\newcommand{\R}{{\mathbb R}}

\newcommand{\C}{{\mathbb C}}
\newcommand{\Z}{{\mathbb Z}}
\newcommand{\cH}{{\cal H}}
\newcommand{\cC}{{\cal C}}
\newcommand{\cD}{{\cal D}}
\newcommand{\cQ}{{\cal Q}}
\newcommand{\cM}{{\cal M}}
\newcommand{\cA}{{\cal A}}

\newtheorem{Theorem}{Theorem}
\newtheorem{Lemma}{Lemma}

\date{August 20, 2007}

\begin{document}

\title{How much is a quantum controller controlled\\ by the controlled system?}

\author{Dominik Janzing\thanks{School of Computer Science and Electrical Engineering, University of Central Florida, Orlando, FL 32816, USA. Electronic address: \texttt{janzing@ira.uka.de}}\,\, and Thomas Decker\thanks{Department of Computer Science \& Engineering, University of
Washington, Seattle, WA 98195, USA. Electronic address: \texttt{decker@ira.uka.de}}}

\maketitle

\abstract{We consider unitary transformations on a bipartite system $A\times B$. 
To what extent entails the ability to transmit information from $A$ to $B$ 
the ability to transfer information in the converse direction?
We prove a dimension-dependent lower bound on the
classical channel capacity $C(A\leftarrow B)$ in terms of the capacity $C(A\rightarrow B)$ for the case that
the bipartite unitary operation consists of controlled local unitaries on $B$ conditioned on basis states on $A$. 
This can be interpreted as a statement on the strength 
of the inevitable backaction of a quantum system on its controller. 

If the local operations are given by the regular representation of a finite group $G$ we have $C(A\rightarrow B)=\log |G|$ and 
$C(A\leftarrow B)=\log N$ where 
$N$ is the sum over the degrees of all
inequivalent representations. 
Hence the information deficit $C(A\rightarrow B)-C(A\leftarrow B)$
between the forward and the backward capacity 
  depends on the ``non-abelianness'' of the control group. For regular representations, the ratio between backward and forward
capacities cannot be smaller than $1/2$. The symmetric group $S_n$ reaches this bound asymptotically. However, for the
general case (without group structure) all bounds must depend on the dimensions since it is known that 
the ratio can tend to zero.
}

\section{Introduction and formal setting}

The fact that no measurement can extract information about a quantum system without disturbing its 
state is one of the essential features of quantum theory \cite{Omnes,Ja68}. 
Roughly speaking,  it is therefore true that the measurement apparatus
influences always the quantum system when the system influences the measurement apparatus. 
For this reason, bidirectionality  
of causal influences seems to be a general feature of quantum theory. 
However, measurement-disturbance relations \cite{Fu96b} for quantum measurements do not provide a general 
answer to the following question:
to what extent is the state of the measured system influenced by the {\it state} of the measurement apparatus?
This question refers to a stronger sense of causal bidirectionality: Whenever an interaction between two systems $A$ and $B$ allows us to transmit information from $A$ to $B$ then it also enables information transfer from $B$ to $A$. In this article we show
that (1) such a stronger sense of causal bidirectionality is true for interactions between {\it finite} dimensional quantum systems
but violated in infinite dimensions and (2) there is a dimension-dependent lower bound on the channel capacity from $B$ to $A$ in terms of the capacity from $A$ to $B$. 

The motivating example to study quantitative relations between the information that can be transmitted from $A$ to $B$ 
and the amount of information that can be sent in the converse direction is the well-known symmetry of the controlled-not gate (``CNOT'') \cite{NC}.
Let $U$ be a CNOT gate with qubit $A$ as {\it control} wire and qubit $B$ as {\it target} wire. This gate allows us to transmit the classical information $1$ bit from $A$ to $B$: To this end, we initialize $B$ to the basis state $|0\rangle$ and choose one of the states $|0\rangle, |1\rangle$  
for system $A$. After applying CNOT to the joint system the state of $B$ will be $|0\rangle$ or $|1\rangle$ depending on which state we have chosen
for $A$. Since the roles of control wire and target wire are swapped when the CNOT gate is described in the Hadamard basis
we can also transmit $1$ classical bit of information from $B$ to $A$ after we have initialized $A$ to the state 
$|+\rangle:=(|0\rangle +|1\rangle)/\sqrt{2}$. 
A possible generalization of this symmetry is the following observation. Let $P_j:=|j\rangle\langle j|$ for $j=0,\dots,n-1$ be the projector onto the span of the
$j$th canonical basis vector in $\C^n$ and $S$ be the cyclic shift operator on $\C^n$ defined by
\[
S:=\sum_{j=0}^{n-1}|j\rangle \langle( j+1)\, {\rm mod}\,\, n|\,.
\]
Then we introduce controlled powers of the shift by
\begin{equation}\label{shiftsym}
U:=\sum_{j=0}^{n-1} P_j\otimes S^j\,.
\end{equation}
Elementary algebra shows that a conjugation of $U$ with a Fourier transform \cite{NC} on both components leads to
\[
\tilde{U}:=\sum_{j=0}^{n-1} S^{-j} \otimes P_j\,.
\]
Since $U$ allows us to send the information $\log_2 n$ bits from $A$ to $B$ we can also transfer $\log_2 n$ bits in the converse
direction by initializing $A$ in one of the Fourier transformed basis states.  
Because this symmetry does not apply to general unitary transformations $U$ we want to understand how to quantify the amount of
information that can be transferred backwards in the general case. 

The motivation to ask this type of questions is given by the following background: 
\begin{itemize} 
\item {\bf Understanding causality.}
The statement that two physical systems {\it interact} defines, in the first place, a symmetric relation. On the other hand, it is a matter of fact that there are situations where the effect of one physical system on a second one is more relevant than the effect of the latter on the former.
To understand under what conditions causal unidirectionality emerges in a way that is consistent with Hilbert space quantum mechanics
would be another small step towards a deeper understanding of the physics of causal directions. 

\item {\bf Understanding fundamental limits of quantum control.}
A quantum controller is a device that influences a quantum system in a desired and flexible way. Quantum  control is often phrased in terms of 
time-dependent Hamiltonians \cite{Ll97,KGB3}, taken from one parameterized set of Hamiltonians.
This description refers to the controller as a classical system. 
 Even though this perspective
is very helpful for practical purposes, 
the following fundamental point of view
may be more helpful to understand the limits and the thermodynamics of quantum control: 
actually, the interaction between controller and system induces  a joint dynamics of the bipartite 
quantum system (which may, in addition, also involve the environment as a third system).
Surprisingly, this perspective has rarely \cite{JZAB,Landahl} been discussed in the context 
of quantum  control even though it was quite popular in the context of quantum measurements \cite{He72}. 
As noted in \cite{JZAB} the arbitrariness of the so-called ``Heisenberg cut'' between the system to be measured and
the measurement apparatus occurs also in the quantum control setting: The question ``who controls the quantum controller?'' could lead 
to a never ending sequence of ``meta-controllers'' and consistency of quantum theory requires that we can shift the cut between the controlled system and its controller.
Toy models for a consistent shift of this kind have been described in \cite{JZAB}.
When asking which feature makes a quantum system interacting with another system the controller of the latter, 
it is natural to explore to what extent the controller is immune to changes of the state of the former.
We do not claim that this immunity is a necessary or sufficient feature of a quantum controller. Nevertheless, 
we are convinced that the thermodynamic limits of quantum control are related to the question which amount of information
is transferred to the controller.

\item {\bf Generalizations of the phase kick-back}.
The symmetry of CNOT, or, more general, the symmetry of the controlled powers of the cyclic shift is just an instance of the well-known phase kick-back
that is used in quantum phase estimation \cite{ClevePhase}. 
It has been shown  that every quantum algorithm can be rewritten in such a way that
it contains phase estimation as its central part \cite{WZ:06}. For this reason, it is desirable to understand in which sense
there are generalizations of the phase kick-back  to non-abelian groups. 
The group structure is actually of minor relevance for the above philosophy-focused questions. However,
representation theory of finite groups will provide us with nice examples where the backaction can be analyzed.

\item{\bf Limits of classical concepts of low power computing}. Due to progressing miniaturization quantum effects are expected to play a dominant role in future 
computing devices. A characteristic feature of current technology is the well-defined direction of the information flow: The input of a device is supposed to control the output, not vice versa. Likewise, the clock signal is supposed to trigger logical operations and not the other way round. 
To what extent an unidirectionality of this kind is  possible if the complete dynamics of the computation process is dominated by
quantum uncertainties is an open question. Limits of this kind have, for instance, been discussed in \cite{Steudel,SynchrEntropy,viva2002,clock}.

\end{itemize}

To address the above questions we
describe the quantum systems  $A$ and $B$ by
Hilbert spaces $\cH_A$ and $\cH_B$, respectively, and consider a unitary operation $U$ on $\cH_A\otimes \cH_B$. 
Assume both systems are independently initialized into quantum states $\rho_A$ and  $\rho_B$.
For every state $\rho_B$ we obtain a channel 
\[
G_{(A\rightarrow B)}(\rho_A):={\rm tr}_A(U (\rho_A\otimes \rho_B) U^\dagger)
\]
and a $\rho_A$-depedent ``backwards'' channel
\[
G_{(A\leftarrow B)}(\rho_B):={\rm tr}_B(U (\rho_A\otimes \rho_B)U^\dagger)\,.
\]
Note that we have dropped the depedence of $\rho_B$ and $\rho_A$, respectively, in our notation.
We define the forward channel capacity by the maximal amount of Holevo information \cite{NC} that can be sent from $A$ to $B$:
\[
C(A\rightarrow B):=\sup \Big\{  S\Big(G_{A\rightarrow B}\Big(\sum_j p_j \gamma_j\Big)\Big)-\sum_j p_j S\Big(G_{A\rightarrow B}(\gamma_j)\Big)\Big\}\,,
\]
where the supremum is taken over all ensembles $\{p_j,\gamma_j\}$ of density operators acting on
$\cH_A$ and all possible initializations $\rho_B$. Here $S$ denotes the von-Neumann entropy.
This capacity has been called Holevo-Schumacher-Westmoreland capacity in \cite{Cortese}. It has been shown to be the
maximal amount of classical information  that can be sent when multiple copies of the channel are available and the 
receiver is able to perform arbitrary joint measurements on the joint output state \cite{HolevoHSW}.
The motivation to  focus on the {\it classical} information capacity rather than on the {\it quantum} capacity 
is that the ability to transfer classical information is already a clear indication for $B$ {\it influencing} $A$. 
The backward capacity is defined in an analogous way. In terms of these definitions, the goal of this paper is to
understand under which circumstances $C(A\leftarrow B)$ can be small even though $C(A\rightarrow B)$ is large.
Bipartite unitary gates as communcation resources have, for instance, been studied in \cite{BennettHarrow,HarrowShor}. 
The major part of the literature that appeared in this context focuses on the capabililities of creating entanglement
\cite{LindenDisEnt,Chefles,ZanardiEnt,ZZF00,CDK01,DuerEnt,DuerEnt2,KrausCirac,LeiferHenderson}, but studies also 
the relation to classical information capacities \cite{BerryS} for the special case of two-qubit systems.
However, a profound understanding of these relations
and tight bounds on backward capacities in terms of forward capacities 
 in arbitrary dimensions are still missing.

\section{Qualitative statements}

We have emphasized that the questions of this article are not answered by the known information-disturbance
relations in any obvious sense.
The following observation makes this difference more apparent: if an interaction transmits 
information about an unknown quantum state of $A$ to
 $B$ then it changes necessarily the state of $A$. This holds regardless of the Hilbert space dimensions 
of $A$ and $B$. However, in infinite dimensions, the way how the interaction changes the state of $A$ can be completely 
independent of the state of $B$.  In other words, in infinite dimensions we may have forward information transmission without backward
information transmission even though measurement-disturbance relations remain valid:

\begin{Lemma}
There exists unitary operations acting on two quantum systems with separable infinite dimensional Hilbert spaces 
such that $C(A\rightarrow B)\neq 0$ but 
$C(A\leftarrow B)=0$.
\end{Lemma}

\noindent
Proof: Let $\cH_A$ and $\cH_B$ be spanned by basis vectors labeled by the binary sequences 
\[
(a_{n})_{n=0,-1,-2,-3,\dots} \quad \hbox{ and } \quad (b_n)_{n=1,2,3,\dots}  \quad  \hbox{ with } \quad a_n,b_n \in \{0,1\}
\] 
respectively, each sequence $(a_n)_n$ and $(b_n)_n$ 
containing finitely many symbols $1$. 
The tensor product $\cH_A\otimes \cH_B$ can be canonically identified with a space whose basis vectors are labeled by
the binary sequences $(c_n)_{n\in \Z}$ having finite Hamming weight since $\cH_B$ corresponds to the positive numbers and 
$\cH_A$ to the negative numbers and $0$.  We can think of the system as an infinite chain of quantum bits (``qubits'') with the additional
restriction that only a finite set of qubits are in its upper state.
A right shift of basis vectors induced by the right shift on $\Z$ is given by
\[
(c_n)_{n\in \Z} \mapsto (c_{n-1})_{n\in \Z}\,,
\]
and will clearly allow us to send one bit from $A$ to $B$ because the state of the rightmost qubit of $A$ is shifted to $B$. 
Nevertheless, the state of $A$ is completely immune with
respect to changing the state of $B$ before the shift has been applied because the final state of $A$ is simply given by shifting the
state of the chain that corresponds to the values $n\leq -1$ one site to the right.
$\square$
\vspace{0.3cm}

\noindent
However, in finite dimensions we have \cite{BennettHarrow}:

\begin{Theorem}
Let $\cH_A$ be finite dimensional.
If for some unitary $U$ on $\cH_A\otimes \cH_B$ the backward channel capacity satisfies $C(A\leftarrow B)= 0$ then also $C(A\rightarrow B)= 0$.  
\end{Theorem}

We give an alternative proof that is purely algebraic and makes apparent that finite dimensionality is only needed for $A$:

\noindent
Proof: 
Assume $C(A\leftarrow B)=0$. 
Then  there is no observable $A$ on $\cH_A$ whose expected value changes if we apply a unitary transformation ${\bf 1}\otimes V$ to a state
$\rho_A \otimes \rho_B$. Hence we have
\[
{\rm tr}( ({\bf 1}\otimes V) U(A\otimes {\bf 1}) U^\dagger ({\bf 1}\otimes V^\dagger) \rho_A \otimes \rho_B)={\rm tr}(U(A\otimes {\bf 1})U^\dagger \rho_A\otimes \rho_B) \,.
\]
 Since this statement holds for all $\rho_A, \rho_B$ we conclude
\[
({\bf 1}\otimes V) U (A\otimes {\bf 1}) U^\dagger ({\bf 1}\otimes V^\dagger) = U(A \otimes {\bf 1})U^\dagger
\]
for all unitary operations  $V$ on $\cH_B$. 
Hence $({\bf 1}\otimes V)$ commutes with all $U(A\otimes {\bf 1})U^\dagger$ for all $V,A$. 
Let $\cM_A$ and $\cM_B$ be the algebra of operators on $\cH_A$ and $\cH_B$, respectively.
The commutant of  the algebra ${\bf 1}\otimes \cM_B$
is given by $\cM_A\otimes {\bf 1}$. Hence $U(\cM_A\otimes {\bf 1})U^\dagger \subset
\cM_A\otimes  {\bf 1}$. For this reason, the conjugation with $U$ defines an injective $C^*$-homomorphisms (see e.g. \cite{Mu90}) 
$\cM_A\otimes {\bf 1}\rightarrow \cM_A \otimes {\bf 1}$. Since $\cH_A$ is finite dimensional it is also
surjective and hence a $C^*$-automorphism. For matrix algebras, every such automorphism is inner \cite{Mu90}, i.e.,
given by conjugation with one of its unitary elements.   

Hence 
there is some $W\in \cM_A$ such that
$U(A\otimes {\bf 1}) U^\dagger=W A W^\dagger\otimes {\bf 1}$ for all $A\in \cM_A$. 
This implies that $(W^\dagger \otimes {\bf 1}) U$ commutes with $\cM_A\otimes {\bf 1}$ and is therefore an element of
${\bf 1}\otimes \cM_B$. Hence  
$U$ 
has the form $U=W\otimes Y$ for some unitary operators $W,Y$. 
This is certainly a symmetric statement with respect to swapping the systems $A$ and $B$.$\square$
\vspace{0.3cm}

\section{Generalizing the CNOT symmetry}

It would be interesting to know the class of unitary transformations for which 
$C(A\leftarrow B)=C(A\rightarrow B)$. 
This equality is, for instance, true for every $U$ acting on $\C^2\otimes \C^2$ because 
$U$ can be decomposed  \cite{KBG2} as
\begin{equation}\label{2dim}
U=(W_A \otimes W_B) \exp\Big( i\sum_{\alpha=x,y,z}c _\alpha \sigma_\alpha \otimes \sigma_\alpha\Big)  (V_A\otimes V_B) \quad \hbox{with}
\quad c_\alpha \in \R\,.
\end{equation}
Since the local unitaries $W_A,W_B,V_A,V_B$ 
are irrelevant, $U$ can be simplified to an operator that is symmetric in $A$ and $B$.
However, in view of the philosophical questions raised in the introduction, statements that refer to particular dimensions are
only of minor interest.
The following set of bipartite unitaries defines a significant generalization compared
to the ones given by conjugating the operator $U$ in  Eq.~(\ref{shiftsym})  with local unitaries on both components:

\begin{Theorem}\label{DiagSym}
Let $U$ be a unitary on $\C^n\otimes \C^m$ of the form
\[
U=(V_A\otimes V_B) D (W_A\otimes W_B)\,,
\]
where $D$ is diagonal in some product basis and $V_A,V_B,W_A,W_B$ are local unitaries.
Then $C(A\rightarrow B)=C(A\leftarrow B)$.
\end{Theorem}

\noindent
Proof: Assume 
\[
U=D=\sum_{i=0}^{n-1} \sum_{j=0}^{m-1} d_{ij} |i\rangle \otimes |j\rangle
\]
 without loss of generality. 
It is clear that the optimal amount of information transfer  can be achieved with pure states. It is furthermore obvious that
the optimum from $A$ to $B$ can be achieved using basis states on $A$. This is because superpositions of basis states
 lead to mixtures of the corresponding output states on $B$. 
Assume we choose basis state $|i\rangle$ with probability $p_i$ and we have $B$ initialized to the state
\[
|\psi\rangle=\sum_{j=0}^{m-1} c_j |j\rangle\,.
\]
Given that the state $|i\rangle$ has been chosen for  $A$ we obtain for $B$ the pure state
\[
|\phi_i\rangle:=\sum_{j=0}^{m-1} d_{ij} c_j |j\rangle\,.
\]
Since the output states are pure the 
Holevo information transferred to $B$ is  given by the von-Neumann entropy of the mixture of outputs, i.e., by
\[
S(\gamma) \quad \hbox{ with } \quad \gamma:=\sum_{i=0}^{n-1} p_i |\phi_i\rangle \langle \phi_i|\,.
\]
We introduce the matrix 
\[
\Phi:=\Big(|\phi_0\rangle, |\phi_1\rangle, \dots,|\phi_{n-1}\rangle \Big)\,,
\]
and rewrite $\Phi$ as the product $\cC\cD$ where $\cD$ denotes the $n\times m$ matrix with entries $d_{ji}$ and $\cC$ is defined by
\[
\cC:={\rm diag}(c_0,\dots,c_{m-1})\,.
\]
Then we can  write $\gamma$ as
\[
\gamma= \cC\cD \cQ\cQ^\dagger \cD^\dagger \cC^\dagger\,,
\]
where 
\[
\cQ:={\rm diag}(\sqrt{p_0},\dots,\sqrt{p_{n-1}})\,.
\]
Since for any two matrices $M$ the spectra of $MM^\dagger$ and $M^\dagger M$ coincide the spectrum of $\gamma$ coincides
with the spectrum of
\[
\cQ^\dagger\cD^\dagger \cC^\dagger \cC\cD \cQ\,.
\]
This is exactly the density matrix we obtain if we choose  the basis states $|j\rangle$ on $B$ with probability $|c_j|^2$ and prepare
$A$ in the state 
\[
\sum_{i=0}^{n-1} \sqrt{p}_i |i\rangle\,.
\] 
This shows that for every protocol that sends basis states from $A$ to $B$  we can construct a scenario to transmit the same amount of information from $B$ to $A$.$\square$

\section{Lower bound on the backward capacity for controlled operations}

\label{BoundSec}
In the remaining part of the paper we will restrict our attention to unitary operators that are unitaries on $B$ controlled by
basis states of $A$. Using the projections $P_j=|j\rangle \langle j|$ for $j=0,\dots,n-1$ we define $U$ on $\C^n\otimes \C^m$ by
\begin{equation}\label{condV}
U:=\sum_{j=0}^{n-1}P_j\otimes V_j\,,
\end{equation}
where each $V_j$ acts on $\C^m$. If $A$ is initialized to the state 
\[
|\phi\rangle =\sum_{j=0}^{n-1} c_j |j\rangle\,,
\]
the input state $|\psi\rangle$ on $B$ leads to the state
\begin{equation}\label{Channel}
G_{A\leftarrow B}(|\psi\rangle\langle \psi|)=\sum_{i,j=0}^{n-1} \overline{c}_i c_j|i\rangle \langle j| \,\langle \psi |V_i^\dagger V_j|\psi \rangle\,.
\end{equation}
We obtain a lower bound on $C(A\leftarrow B)$ in terms of a quantity that measures how much
the operators $V_j$ differ with respect to the operator norm:

\begin{Theorem}\label{Ivond}
Given a bipartite unitary operation
of the form (\ref{condV}).
Let
\begin{equation}\label{dDef}
d:=\max_{j,k} \min_\phi \| V_j-V_k e^{i\phi}\|
\end{equation}
be the maximal distance between the transformations $V_j$.
Then 
\begin{equation}\label{Upper}
C(A\leftarrow B) \geq H_2\Big(\frac{1}{2}+\frac{\sqrt{1-d^2/4}}{2}\Big)\,, 
\end{equation}
where $H_2(x):=-x\log_2(x) -(1-x)\log_2(1-x)$ denotes the binary entropy function.
\end{Theorem}

\noindent
Proof: Let $A$ be initialized to the state
\[
|\phi\rangle:=\frac{1}{\sqrt{2}} (|j\rangle +|k\rangle)\,,
\]
where $j,k$ denote the pair maximizing expression~(\ref{dDef}). Let $|\psi_1\rangle,|\psi_2\rangle$ be eigenstates of 
$U_jU_k^\dagger$ with eigenvalues $e^{i\mu_1}$ and $e^{i\mu_2}$ such that $|e^{i\mu_1}-e^{i\mu_2}|=d$. Then 
\[
\ell:=\frac{1}{2}(e^{i\mu_1} +e^{i\mu_2})  
\]
has the absolute value $|\ell|=\sqrt{1-d^2/4}$.
If one chooses one of the states $|\psi_p\rangle$ with $p=1,2$ it follows from Eq.~(\ref{Channel}) that the output state is given by
a pure state. It is  supported by the two-dimensional space spanned by $|j\rangle$ and $|k\rangle$ and reads: 
\begin{equation}
\sigma_p:=\frac{1}{2}\left(\begin{array}{cc}  1& e^{i\mu_p}\\ e^{-i\mu_p} & 1 \end{array}\right)\,.
\end{equation}
The uniform mixture 
\[
\frac{1}{2} (\sigma_1 +\sigma_2)\,
\]
has the off-diagonal entries $\ell/2$ and $\overline{\ell}/2$ and thus the eigenvalues $1/2\pm |\ell|/2$.  Its entropy is therefore given by
$H_2(1/2+ |\ell|/2)$. $\square$
\vspace{0.3cm}

\noindent
Given the maximal distance $d$ we can derive a dimension-dependent upper bound on the forward channel capacity:

\begin{Lemma}\label{dVonI}
Let $V_j$ be a set of unitaries with a given maximal distance $d$. Then 
\[
C(A\rightarrow B) \leq  \min\Big\{ \log k,\,\,  H_2(d/2) + \frac{d}{2} \log (k-1) \Big\}\,,
\]
where $k$ is the minimum of $n$ and $m$.
\end{Lemma}

\noindent
Proof: Let $|\psi\rangle$ be the state of $B$.
If one chooses the $j$th basis state of $A$ with probability $p_j$ one obtains on $B$ the state 
$V_j |\psi\rangle \langle \psi |V_j^\dagger$ with probability $p_j$. The entropy of the mixture
\[
\sigma:=\sum_j p_j V_j|\psi\rangle \langle \psi | V_j^\dagger
\]
coincides with $C(A\rightarrow B)$ if the optimal pair $p$ and $|\psi\rangle$ have been chosen.
The entropy can be bounded from above as follows.
Due to 
\[
\|V_j|\psi\rangle \langle \psi |V^\dagger_j -V_0|\psi\rangle \langle \psi|V^\dagger_0\|_1\leq d
\]
and the convexity of the trace norm we have 
\begin{equation}\label{Tr}
\|\sigma-V_0|\psi\rangle \langle \psi|V^\dagger_0 \|_1 \leq d\,.
\end{equation}
The rank of $\sigma$ is at most $k:=\min\{n,m\}$.
Let $q_1,\dots,q_k$ be the diagonal entries of $\sigma$ with respect to a basis
of the image of $\sigma$ 
 that  contains $V_0|\psi\rangle$
as its first basis vector. We derive an upper bound on the probability distribution $q$ which is also an upper bound
on the von Neumann entropy of $\sigma$. 
The diagonal entries of $V_0|\psi\rangle \langle \psi|V^\dagger_0$ are
$1,0,\dots,0$ and the trace-norm distance on the left hand side of Eq.~(\ref{Tr}) is at least $2\sum_{j\geq 2} q_j$. 
This implies $s:=\sum_{j\geq 2} q_j \leq d/2$.  If $d/2\geq (k-1)/k$ 
then $q$ could even be the uniform distribution and we obtain only the trivial bound $S(\sigma) \leq \log k$. 
Otherwise, we obtain maximal Shannon entropy for $q$ if we distribute $s$ uniformly 
on the indices $2,\dots,k$ which is the distribution
$1-d/(2k-2),d/(2k-2),d/(2k-2),\dots,d/(2k-2)$. 
Its Shannon entropy  
is $H_2(d/2)+(d/2) \log (k-1)$. 
$\square$

\vspace{0.3cm}
\noindent
The right hand side of Ineq.~(\ref{dVonI}) is a strictly monotonic function $d\mapsto f(d)$ for $d\leq 2(k-1)/k$. 
In this regime we have therefore the bound
\[
d \geq f^{-1} \left(C(A\rightarrow B)\right)\,.
\]
By inserting the right hand side into  Ineq.~(\ref{Upper}) we obtain an explicit lower bound on $C(A\leftarrow B)$ in terms
of $C(A\rightarrow B)$. 

The ratio between backward and forward capacity allowed by this bound gets small for high dimensions. But this has to be the case
because there is a gate \cite{HarrowShor} in dimension $n\times n$ for which the forward capacity is $\log n$ and the backward capacity
is $O(\log \log n)$. The gate is of the form $U=P_j\otimes V_j$ with
\begin{eqnarray*}
V_j|0\rangle &=&|j\rangle  \\
V_j|i\rangle&=&|i-1\rangle \hbox{ for } 0<i\leq j,  \quad \hbox{ and} \quad V_j |i\rangle=|i\rangle \hbox{ for } i>j\,.
\end{eqnarray*}
Tight bounds on the ratio $C(A\leftarrow B) /C(A\rightarrow B)$ are, however, not known.

\section{Regular representations of finite groups}

\label{Sec:Reg}
Now we restrict our attention to the case where $U=\sum_j P_j \otimes V_j$ 
acts on systems with equal dimension $n$. The extreme 
case  $C(A\rightarrow B)=\log n$ is of course of special interest. Then 
the $V_j$ are sufficiently different  to generate mutually orthogonal states from a given one.
The following construction provides a family of unitaries that satisfy this condition and have enough structure to  
allow us a systematic analysis. Even though this construction does not describe any real physical system, it is nevertheless
helpful because the 
goal of this paper is to explore limitations on the relation between action and backaction that follow from Hilbert space geometry alone
without any specific physical assumptions. 

Let $G$ be a group with $|G|=n$ elements and $(V_g)_{g\in G}$ be the permutation matrices corresponding to
the regular representation of $G$. We label the basis states of $\C^n$ by the elements $g\in G$ and denote them by
$|g\rangle$.  We define
\begin{equation}\label{Reg}
U:=\sum_{g\in G} P_g\otimes V_g\,.
\end{equation}
By initializing $B$ to the state $|{\bf 1}\rangle$, where ${\bf 1}\in G$ denotes the identity element, we clearly can 
obtain $n$ mutually orthogonal states $|g\rangle$ in $B$ by choosing the states $|g\rangle$  with $g\in G$ as inputs, i.e., 
$C(A\rightarrow B)=\log |G|$.
Then we have:

\begin{Lemma}
Let $A$ be initialized to the uniform superposition $\sum_g |g\rangle /\sqrt{|G|}$. Then the set of possible
output states is given by the set of positive matrices with trace one contained in $R\cA R$, 
where $R$ is the reflection $|g\rangle \mapsto |g^{-1}\rangle$ 
and $\cA$ is the 
$C^*$-algebra generated by the matrices $V_g$.
\end{Lemma}

\noindent
Proof: 
 Let
\[
|\psi\rangle:=\sum_g c_g |g\rangle 
\]
be an arbitrary input state. Due to Eq.~(\ref{Channel}) the output state $\sigma$ on $A$ is given by
\[
\Big(\frac{1}{|G|}\sum_{gh} \overline{c}_g c_h \langle g | V_{m^{-1}r} |h\rangle\Big)_{m,r\in G} \,.
\]
The inner product is $1$ for all $mg=rh$ and $0$ otherwise.
Elementary calculations show
\[
\sigma=R\left(\frac{1}{|G|}\left(\sum_g \overline{c}_g V_g\right) \left(\sum_h c_h V^\dagger_h\right)\right)R\,.
\]
Thus, $\sigma$ is, up to the inversion $R$, an element of $\cA$
(which is isomorphic to the group algebra $CG$ \cite{ClausenBaum}).
Let $\sigma$ be an arbitrary positive element of  $R\cA R$ with trace one. Since $R\cA R$ is closed with respect to square roots we can find
an $a\in R\cA R$ such that $a^\dagger a=\sigma$. We can $a$ write as
$a=R\sum_g c_g V_gR/\sqrt{|G|}$  and hence we obtain $\rho=R(\sum_g c_g V_g)^\dagger (\sum_g c_g V_g)R/|G|$. 
Due to $\sum_g |c_g|^2={\rm tr}(\rho)=1$ the coefficient vector $(c_g)_{g\in G}$ is  a unit vector and represents therefore a possible input state.$\square$
\vspace{0.3cm}

Using the explicit characterization of output states for the case that $A$ is initialized to a uniform superposition
we can calculate the backward channel capacity (even without restricting to uniform initializations):
\begin{Theorem}
The backward information capacity satisfies
\[
C(A\leftarrow B) = \log N\,,
\]
where $N$ is the sum of the degrees of all inequivalent irreducible representations of $G$.
\end{Theorem}

\noindent
Proof: 
Let us first assume that $A$ is initialized to $\sum_g |g\rangle /\sqrt{|G|}$.
The algebra $\cA$ generated by the representation matrices $V_g$ is given by
 \cite{ClausenBaum,Serre}
\begin{equation}
\cA=F^\dagger \Big( \bigoplus_r {\bf 1}_r\otimes \cM_r  \Big) F\,.
\end{equation}
The sum runs over all inequivalent representations $r$. Their degree is denoted by $d_r$
and  ${\bf 1}_r$ denotes the identity of dimension $d_r$.
 $F$ is the generalized Fourier transform that
achieves block diagonalization of $\cA$.  
The multiplicities are irrelevant for the
channel capacity. Therefore we may identify the set of possible output states
with the density matrices in
\[
\bigoplus_r M_r\,,
\]
acting on a Hilbert space of dimension $N=\sum_r d_r$. This shows that the capacity is at most $\log N$.
On the other hand, we can obtain every output state that is given by
one entry $1$ on one of the $N$ diagonal positions. Hence, the capacity is $\log N$. 
 
Now we drop the assumption that $A$ is initially in a uniform superposition. Instead, we assume
\[
|\phi\rangle:=\sum_g \sqrt{p_g}|g\rangle
\]
to be the state of $A$ before $U$ is applied.
With respect to the original basis $|g\rangle$,
this changes the output according to the map
\begin{equation}\label{DDeform}
F_p:\sigma\mapsto D\sigma D\,.
\end{equation}
where $D$ is the diagonal operator with entries $\sqrt{p_g} \sqrt{|G|}$.
The domain of $F_p$ is the smallest $C^*$-algebra containing every possible output state, i.e.,
$R\cA R$. 
Note that this specification of the domain makes $F_p$ trace-preserving because all elements of $R\cA R$ are constant
along the diagonal. This implies that the diagonal of $D\sigma D$ is given by the values $p_g$ if
$\sigma$ has the diagonal entries $1/|G|$.
Hence we have shown that the deformation is a quantum channel. Thus, the deformed outputs cannot provide more information 
about the input than the undeformed outputs by monotonicity of Holevo information \cite{PetzMon}.
$\square$

\vspace{0.3cm}
\noindent
The number $N$ coincides with $|G|$ if and only if $G$ is abelian. This is because the number of inequivalent representations
coincides with the number of conjugacy classes
 \cite{Serre} which is $|G|$ for abelian groups. For non-abelian groups, we have necessarily representations of degree greater than $1$ and 
hence multiplicities greater than $1$. 
This leads immediately to 
the following observation:
\vspace{0.2cm}

\noindent
{\bf Corollary:}\,\,{\it  If $A$ controls the regular representation matrices of  $G$ on $B$ then 
\[
C(A\rightarrow B)=C(A\leftarrow B)
\]
if and only if $G$ is abelian.
}
\vspace{0.3cm}

Even though it is not known what the smallest posisble ratio would be 
for $C(A\leftarrow B)/C(A\rightarrow B)$ representation theory provides a lower bound for the case of regular representations:

\begin{Lemma}\label{Martin}
Let $g\mapsto V_g$ be the regular representation of a finite group. Then the ratio between backward and forward capacity satisfies
\[
\frac{C(A \leftarrow B)}{C(A\rightarrow B)} \geq \frac{1}{2}\,.
\]
\end{Lemma}

\noindent
Proof: We have
\[
N^2 =\left(\sum_r d_r\right)^2 \geq \sum_r d^2_r =|G|\,.
\]
Hence $\log N \geq (\log G)/2$.$\square$.

\begin{figure}
\centerline{\includegraphics[scale=0.7]{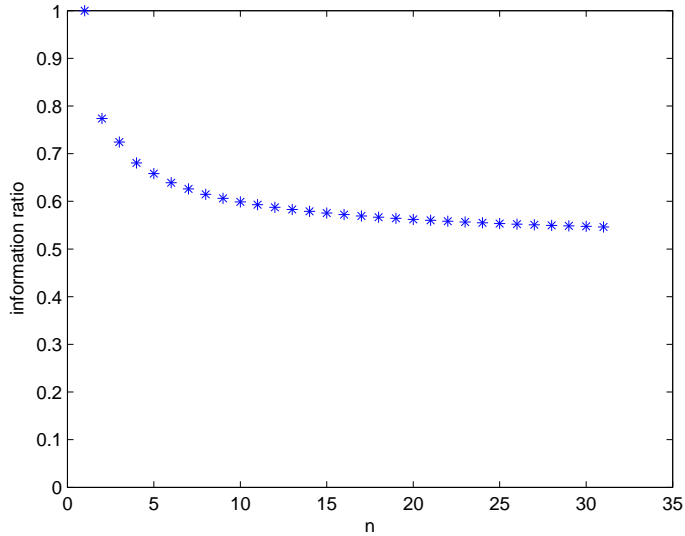}}
\caption{\label{Dia}{\small Ratio $C(A\leftarrow B)/C(A\rightarrow B)$ for the regular representation of $S_n$ for $n=2,\dots,32$.
Note that $S_2$ leads to the controlled-not gate.}}
\end{figure}

\vspace{0.3cm}
\noindent
The lower bound $1/2$ is asymptotically reached by $G=S_n$, i.e., the symmetric group on $n$ points if $n$ tends to infinity. 
With $|G|=n!$ we obtain an estimation of $C(A\rightarrow B)$ from Stirling's formula
stating that $n!$ increases with $\sqrt{2\pi n} (n/e)^n (1+O(1/n))$. If we measure information in terms of natural units, we obtain hence
\[
\lim_{n\to \infty} \Big(C(A\leftarrow B) -(\ln  (\sqrt{2\pi n}) +n\ln n -n )\Big)=0\,.
\] 
An upper bound on $C(A\leftarrow B)$ can be derived from
\[
\sum_r d_r \leq p_n m_n\,,
\]
where $p_n$ is the number of inequivalent representations of $S_n$ and $m_n$ the degree of the largest representation, i.e.,
$m_n:=\max_r \{ d_r\}$. 
An upper bound on $m_n$  is given by \cite{McK}:
\[
m_n \leq (2\pi n)^{1/4} \left(\frac{n}{e}\right)^{n/2}\,.
\]
For $p_n$ we have \cite{Andrews}
\[
p_n \leq c \exp\left(\pi \sqrt{\frac{2}{3} n}\right)\,,
\]
with an appropriate constant $c$.
With $\log N \leq \log p_n +\log m_n$ the only asymptotically relevant term is $(n \ln n)/2$.
For the asymptotics of $\log |G|$, the dominating term is $n\log n$.  Hence $S_n$ reaches asymptotically the minimal possible quotient $1/2$. 
 Figure~\ref{Dia} shows that the values
get already quite close to $1/2$ for $n=32$.

\section{The symmetric group $S_3$}

In this section we want to provide a bit more intuition about the general results of the previous sections. 
The smallest non-abelian group  is $S_3$, the set of permutations of $3$ elements (which is isomorphic to the 
dihedral group $D_3$). Since $|S_3|=3!=6$ the unitary operation defined by the regular representation according to Eq.~(\ref{Reg})
leads to the bipartite system $\C^6\otimes \C^6$. We choose the generating transpositions $a:=(1\,2)$ and $b:=(2\,3)$.
In every component $\C^6$ the basis vectors are labelled by $g$ where we have chosen  the following order of the elements
\[
[g_1,\ldots, g_6]:=[(),(2\,3),(1\,2),(1\,2\,3),(1\,3\,2),(1\,3)]\,.
\]
The group $S_3$ has three inequivalent representations of dimensions $d_1=1$, $d_2=1$, and $d_3=2$. 
Up to unitary equivalence, they  are given as follows.
\[
\tau_1(g):=(1) 
\quad \hbox{ and } \quad
\tau_2(g):=({\rm sgn}(g)) 
\]
with the signum function ${\rm sgn}$. 
Here $(\cdot)$ denotes a $(1\times 1)$-matrix. 
The two-dimensional representation $\tau_3$ is given by 
\[
\tau_3(a):=\left(\begin{array}{cc}0&\omega_3^2\\ \omega_3 & 0
\end{array}\right) \quad {\rm and} \quad 
\tau_3(b):=\left(\begin{array}{cc}0&\omega_3\\ \omega_3^2& 0
\end{array}\right)
\]
where $\omega_3$ is a third complex root of unity.

$\tau_1$ and $\tau_2$  occur with multiplicity $1$ 
and $\tau_3$ with multiplicity $2$. If system $A$ is in a uniform superposition
the set of possible output states $\sigma$ is unitarily equivalent to the set
\begin{equation}\label{Output3}
(p_1)\oplus (p_2) \oplus  \frac{1}{2} p_3 (\sigma_2 \oplus \sigma_2)\,,
\end{equation}
where the non-negative scalars $p_1,p_2,p_3$ with $\sum_j p_j=1$  
define a probability distribution and $\sigma_2$ is an arbitrary two-dimensional density matrix.

The isomorphism between the possible output states with respect to the original basis $|g\rangle$ is described by the
Fourier matrix $F$ that decomposes the regular representation
\[
\tau_{\rm reg}(g)=\sum_h \ket{gh}\bra{h}
\]
into the direct sum 
\[
F^\dagger \tau_{\rm reg}(g) F = \tau_1(g) \oplus \tau_2(g) \oplus 
\tau_3(g) \oplus \tau_3(g)\,.
\]
For our example we find
the unitary
\[
F=\frac{1}{\sqrt{6}}\left( \begin{array}{cccccc}
1&1&\sqrt{2}&0&0&\sqrt{2}\\
1&-1&0&\sqrt{2}\, \omega_3&\sqrt{2}\, \omega_3^2&0\\
1&-1&0&\sqrt{2}&\sqrt{2}&0\\
1&1&\sqrt{2}\,\omega_3&0&0&\sqrt{2}\,\omega_3^2\\
1&1&\sqrt{2}\,\omega_3^2&0&0&\sqrt{2}\,\omega_3\\
1&-1&0&\sqrt{2}\,\omega_3^2&\sqrt{2}\,\omega_3&0 
\end{array}\right)\,.
\]
We choose the following $4$ input states:
\[
\begin{array}{ccccccc}
\ket{\phi_1}&:=&\frac{1}{\sqrt{6}}(1,1,1,1,1,1)^T\,, &\hspace{1cm}& \ket{\phi_2}&:=&\frac{1}{\sqrt{6}}(1,-1,-1,1,1,-1)^T\\
\ket{\phi_3}&:=&\frac{1}{\sqrt{3}}(1,0,0,\omega_3^2,\omega_3,0)^T, &\hspace{1cm}& \ket{\phi_4}&:=&\frac{1}{\sqrt{3}}(0,1,\omega_3,0,0,\omega_3^2)^T\,.
\end{array}
\]
They generate the output states
\begin{equation}\label{OutputDen}
\begin{array}{ccccccc}
\sigma_1&:=&{\rm diag}(1,0,0,0,0,0), &\hspace{1cm}& \sigma_2&:=&{\rm diag}(0,1,0,0,0,0) \\
\sigma_3&:=&\frac{1}{2} {\rm diag}(0,0,1,0,1,0)\,, &\hspace{1cm}& \sigma_4&:=&\frac{1}{2} {\rm diag}(0,0,0,1,0,1)\,.
\end{array}
\end{equation}
We obtain $C(A\rightarrow B)=\log 6$ and $C(A\leftarrow B)=\log 4$.

The following example shows that the statements of Section~\ref{Sec:Reg} do not apply to non-regular representations.
If we choose the usual permutation representation of $S_3$ we obtain a unitary transformation on $\C^6\otimes \C^3$, where
the action of $S_3$ on $\C^3$ is defined by 
\[
\tau(a)=\left( \begin{array}{ccc}0&1&0\\1&0&0\\0&0&1\end{array}\right)
\quad {\rm and} \quad
\tau(b)=\left( \begin{array}{ccc}1&0&0\\0&0&1\\0&1&0\end{array}\right)\,.
\]
We have $C(A\rightarrow B)=\log 3$ because we can clearly obtain $3$ mutually orthogonal states on $B$ by choosing any canonical basis
vector of $\C^3$ as initial state.
Even though $S_3$ is non-abelian we also have  $C(A\leftarrow B) =\log 3$.
Using the three input vectors
\[
\ket{\Phi_1}:=\frac{1}{\sqrt{3}}
\left(\begin{array}{c}1\\1\\1\end{array}\right)
\hspace{1cm}
\ket{\Phi_2}:=\frac{1}{\sqrt{3}}
\left(\begin{array}{c}1\\ \omega_3 \\ \omega_3^2\end{array}\right) \hspace{1cm}
\ket{\Phi_3}:=\frac{1}{\sqrt{3}}
\left(\begin{array}{c}1\\ \omega_3^2\\ \omega_3\end{array}\right)
\]
the output density operators read $\sigma_1,\sigma_3,\sigma_4$ 
 as defined  in Eq.~(\ref{OutputDen}).

\section{What's the message?}

We have shown that a unitary operation on a bipartite finite-dimensional 
system $A,B$ can only enable information transmission from $A$ to $B$ 
whenever there is also information transmission possible from $B$ to $A$. However, for arbitrarily high dimensions the difference between
backward channel capacity and forward capacity can be arbitrarily large.
To show this
we have constructed bipartite unitary operations 
of the following type: 
 mutually orthogonal states on $A$ control the implementation of unitary operations on $B$ taken from a finite group.
Then the backaction becomes smaller the less abelian the group is.

To link our results to the philosophical questions raised in the introduction we assume that  $A$ is a toy model of a quantum controller.
We assume that the interaction between controller $A$ and the system to be controlled (denoted by 
$B$) implements 
\[
U=\sum_j P_j \otimes V_j\,,
\]
(where $P_j$ may also be degenerate projections)
after influencing the system during some fixed time interval.
Let $|\phi\rangle$ be the state of system $A$.  If $|\phi\rangle \in P_j \cH_A$ for some $j$ 
the system $A$ is insensitive. 
However, if the controller state is switched 
from one subspace $P_j \cH_A$   to another
$P_i\cH_A$ there must be a moment where it is a superposition. During the switching process, $A$ is necessarily influenced by $B$
and this paper has tried to clarify to what extent this influence depends on the state of $B$.
For small dimensions, this back action cannot be arbitrarily small. 
At first glance, dimension dependent bounds seem to be of minor interest if one thinks of the quantum controller as a large quantum system.
However, our bound  in Section~\ref{BoundSec} depends on the {\it minimum} of the dimensions of controller and system.
To find tight lower bounds on the backward capacity in terms of the
foward capacity has to be left to the future. 

\section*{Acknowledgments}
The authors would like to thank Martin R\"{o}tteler and
Pawel Wocjan for helpful discussions and Aram Harrow for useful comments.
TD was supported under ARO/NSA quantum algorithms grant number
W911NSF-06-1-0379.


\end{document}